\documentclass[prb,aps,twocolumn,reprint,showpacs,floatfix]{revtex4-1}
\usepackage[dvipdfmx,hyperindex,breaklinks,colorlinks=true,bookmarks=false,linkcolor=blue,citecolor=blue,urlcolor=blue]{hyperref}
\usepackage{dcolumn}
\usepackage{amssymb,amsmath,bm}
\usepackage{times}
\usepackage{graphicx,color}
\usepackage{ulem}

\begin{document}
\title{Pulse excitation to continuous-wave excitation \\ in 
a low-dimensional interacting quantum system}
\author{Atsushi Ono}
\author{Hiroshi Hashimoto}
\author{Sumio Ishihara}
\affiliation{Department of Physics, Tohoku University, Sendai 980-8578 Japan}

\begin{abstract}
Real-time dynamics in one-dimensional transverse Ising model coupled with the time-dependent oscillating field is analyzed by using the infinite time-evolving block decimation algorithm and the Floquet theory. 
In particular, the transient dynamics induced by the pulse field and their connections to the dynamics by a continuous-wave field are focused on. 
During the pulse-field irradiation, the order parameter shows a characteristic oscillation, in which 
the frequency shifts from the pulse-field frequency. 
This is considered as a kind of the Rabi oscillation, but the frequency strongly depends on the intersite Ising interaction. 
After turning off the pulse field, the oscillation remains with a frequency $\mathit{\Omega}$ and a damping constant $\gamma$. 
In the case of low fluence,
both $\mathit{\Omega}$ and $\gamma$ are scaled by the pulse amplitude in a wide range of the parameter values of the model. 
In the case of high fluence, 
$\mathit{\Omega}$ and $\gamma$ are arranged by a product of the pulse amplitude and the pulse width. 
This implies that the dynamics after turning off the pulse field are decided by a population of the excited state when the pulse field is turned off. 
\end{abstract}
\maketitle

\section{introduction}

Light-induced ultrafast dynamics and functional controls in correlated electron systems are widely recognized as fascinating issues in recent condensed matter physics.~\cite{*[{{\it Special Topics: Photo-Induced Phase Transitions and their Dynamics}, edited by }][{}]Gonokami2006, Shen2014a, Eckstein2013, Lu2015a, Lenarcic2014, Kemper2015, Tsuji2012, Yonemitsu2015, Mentink2015}
Photoinduced Mott-insulator to metal transition by femto-second light pulse has been realized nowadays in several classes of transition-metal oxides,~\cite{Miyano1997, Iwai2003} and organic-molecular compounds.~\cite{Tajima2005, Kawakami2010}
A small number of photons bring about macroscopic changes in electronic structures of correlated electron materials.
A variety of exotic phenomena induced by laser pulse irradiations, e.g. the photo-induced superconductivity,~\cite{Fausti2011, Nicoletti2014} demagnetization,~\cite{Beaurepaire1996, Vomir2005, Cheng2016} spin-state transition~\cite{Ogawa2000, Tayagaki2001, Watanabe2009} and so on, are interpreted as consequences of the strong entanglements between the multiple degrees of freedom under strong electron correlation.

Beyond such ultrafast responses induced by the short light pulse, coherent controls of the correlated electron systems by light have recently attracted much attention.~\cite{Tsuji2012, Yonemitsu2015, Ishikawa2014, Mentink2015, Miyamoto2013, Matsubara2014, Fukaya2015, Hashimoto2016, Matsunaga2014, Kampfrath2011}
These are promoted by the recent developments in the THz light source and related techniques. 
Several collective excitations and elementary excitations, such as phonon, magnon, and orbiton, are able to be excited directly by the THz light.
As a consequence, macroscopic magnetic, electric, and lattice structures are changed coherently. 
In the theoretical side, the coherent responses under the continuous-wave (CW) light are well described by the Floquet theory.~\cite{Shirley1965, Sambe1973} The nonequilibrium dynamics under the time-periodic light are replaced by the time-independent eigen state problem. 
Owing to its technical advantageous and clear physical picture, this formalism is now applied to wide issues in correlated electron systems under the time-periodic external field.~\cite{Grossmann1991, Oka2009, Lindner2011, Kitagawa2011, Tsuji2009, Takayoshi2014b, Eckardt2005, Mori2016}

Instead of such recent developments in experimental and theoretical researches, the coherent light control of correlated electron systems is still under way. 
One of the reasons is attributable to the technical limitations at the present stage in the experiments in which combining the strong electric field and the long-lived light pulse are difficult. 
For example, a THz laser pulse with a few pico-second width oscillates only a few times. 
In the theoretical side, it is not clear whether the standard Floquet theory is applicable to a short pulse or not. 
It is widely required to reveal a connection between the transient dynamics induced by the short pulse and those by the CW light in strongly correlated systems. 

In this paper, the transient dynamics induced by the pulse external field and their connection to the dynamics by the CW field are studied in a low-dimensional interacting quantum system. 
We adopt the one-dimensional transverse Ising (TI) model with the time-dependent external field with 
the frequency $\omega_{\rm p}$, the amplitude $A$, and the pulse width $t_{\rm width}$. 
This model is relevant for several correlated electron systems, such as the dimer-type organic molecular solid,~\cite{Naka2013, Naka2010} and the excitonic insulators.~\cite{Batista2002, Nasu2016}
The situation has some similarities to the studies in interacting cold atom systems.~\cite{Cummings1983, Lukin2001, Dudin2012}
It is stressed that
the present calculation method based on the infinite time-evolving block decimation (iTEBD) algorithm~\cite{Vidal2007} treats the quantum many-body effects as well as the transient dynamics exactly within numerical errors without a finite-size effect.
During the pulse-field irradiation, the order parameter shows a characteristic oscillation, in which the frequency, $\mathit{\Omega}$, shifts from $\omega_{\rm p}$. 
Through the analyses by the Floquet theory, this oscillation is considered as a kind of the Rabi oscillation, but the frequency shift $\mathit{\Omega}-\omega_{\rm p}$ strongly depends on the inter-site Ising interaction.
After turning off the pulse field, this oscillation remains, and its amplitude is damped. 
In the case of low fluence, 
both $\mathit{\Omega}-\omega_{\rm p}$ and the damping constant, $\gamma$, follow the scaling curves in a wide range of the parameter values. 
In the case of high fluence, $\mathit{\Omega}-\omega_{\rm p}$ and $\gamma$ are varied strongly by $A$ and $t_{\rm width}$, but are arranged by a product of the two. 
These results imply that the dynamics after turning off the pulse field are decided by a population of the excited state when the pulse field is turned off.

In Sec.~\ref{sec:model}, the model Hamiltonian and the numerical method are introduced. 
In Secs.~\ref{sec:itebd} and \ref{sec:floquet}, respectively, the results of the numerical simulation in an infinite system, and the analyses by the Floquet theory are presented. 
Section~\ref{sec:discussion} is devoted to summary. 

\section{model and method}
\label{sec:model}
We adopt the TI model on a one-dimensional chain interacting with a time-dependent external field, which is one of the simplest quantum many body model. This is defined by 
\begin{align}
{\cal H} = {\cal H}_{\rm TI} + {\cal H}_V.
\label{eq:htot}
\end{align}
The first term is the standard TI model given by 
\begin{align}
{\cal H}_{\rm TI} = -J \sum_i \sigma_i^x \sigma_{i+1}^x - h_z \sum_i \sigma_i^z , 
\label{eq:ti}
\end{align}
and the the second term represents an interaction with the oscillating external field with finite width given by 
\begin{align}
{\cal H}_V = -h_x(t) \sum_i \sigma_i^x
\label{eq:hv}
\end{align}
with
\begin{align}
h_x(t) = A w(t) \cos(\omega_{\mathrm{p}} t) . 
\label{eq:field} 
\end{align}
The envelope function is chosen to be 
\begin{align}
w(t) = \begin{cases}
e^{-t^2/(2t_{\mathrm{p}}^2)} & (t < 0) \\
1 & (0 \leq t \leq t_{\text{width}}) \\
e^{-(t-t_{\text{width}})^2/(2t_{\mathrm{p}}^2)} & (t_{\text{width}}<t) ,
\end{cases}
\label{eq:field2}
\end{align}
where $t_{\rm width}$ represents the time interval of the external field. 
When $t_{\rm width}=0$, we have 
\begin{align}
w(t)= e^{-t^2/(2t_{\mathrm{p}}^2)}
\label{eq:envelope}
\end{align}
for all $t$. 
We define that $\sigma_i^\alpha\ (\alpha=x,y,z)$ are the Pauli matrices at site $i$, $J$ is the exchange interaction, and $h_z$ is the transverse field. 
The oscillating pulse field coupled to $\sigma^x$ is represented by $h_x(t) $ with amplitude $A$, width $t_{\mathrm{p}}$, and frequency $\omega_{\mathrm{p}}$.
All parameters introduced above are chosen to be positive.
This model is nonintegrable owing to the external field.

As is well known, the ground state of the TI model without the oscillating external field is an ``ordered state" in $J>h_z$ where $m^x \equiv N^{-1} \langle \sum_i \sigma_i^x \rangle$ is finite, a ``disordered state" in $J<h_z$ where $m^x=0$, and the quantum critical point at $J=h_z$.
We denote the number of the sites by $N$. 
The one-dimensional TI model is diagonalized by the Jordan-Wigner transformation~\cite{Pfeuty1970} as 
$ H_{\rm TI} = \sum_k \varepsilon_k \eta_k^\dagger \eta_k + \text{const.} $,
where $\eta_k^\dagger$ and $\eta_k$, respectively, are the creation and annihilation operators of a fermion with momentum $k$, and the $\varepsilon_k$ is the energy dispersion defined as 
$\varepsilon_k = 2\sqrt{J^2 + h_z{}^2 - 2 J h_z \cos k}$. 
The energy gap is identified as ${\mathit{\Delta}} \equiv \varepsilon_{k=0} = 2\lvert h_z - J \rvert$ .

The TI model with the time-dependent external field introduced in Eq.~(\ref{eq:htot}) has a number of implications for the ultrafast optical dynamics in correlated electron systems. 
One system to which the present model is applicable is the quasi one-dimensional organic compounds with molecular dimers, e.g. (TMTTF)$_2$X (X: anion molecule).~\cite{Naka2013, Naka2010}
Molecule pairs are aligned in a quasi-one dimensional chain, 
and two-outermost molecular orbitals in each dimer unit construct the bonding and antibonding bands. 
Since one electron exists per the dimer unit, this is recognized as a Mott insulator when the Coulomb interaction is strong enough in comparison with the band width. 
The electronic state inside of the $i$th dimer is denoted by the spin operators; $\sigma_i^z=+1\ (-1)$ represents the state where the electron occupies the bonding (antibonding) orbital, and $\sigma_i^x=+1\ (-1)$ represents the state where the electron locates in the right (left) molecule, i.e. the directions of the electric-dipole moment inside a dimer. 
The first and second terms in Eq.~(\ref{eq:ti}) correspond to the inter-dimer Coulomb interaction, and the intra-dimer electron hopping, respectively. 
The time-dependent external field in Eq.~(\ref{eq:hv}) describes the interaction between the optical laser pulse and the electric dipole moments inside the dimers. 
Another system to which the present model is applicable is the excitonic insulating systems, where 
the spontaneous wave-function mixing of the valence and conduction bands occurs in the narrow-gap semiconductors and semimetals.~\cite{Batista2002, Nasu2016}
Difference between the electron number densities in the valence and conduction bands, and the mixing between the two bands, which corresponds to the order parameter of the excitonic insulating state, are represented by $\sigma^z$ and $\sigma^x$, respectively. 
The first and second terms in Eq.~(\ref{eq:ti}) correspond to the excition-exciton interaction, and the energy difference between the two bands.
The time-dependent external field in Eq.~(\ref{eq:hv}) describes the interaction between the laser pulse and the excitons. 

The ground states and the time-evolved states after the external-field pumping are analyzed numerically using the iTEBD algorithm.~\cite{Vidal2007} 
The wave function is represented as the infinite matrix-product state as 
\begin{align}
|\Psi \rangle=\sum_{\sigma_1, \sigma_2 ,\dots} \mathop{\textrm{Tr}} (A^{\sigma_1} A^{\sigma_2} \cdots ) 
|\sigma_1, \sigma_2, \dots \rangle , 
\end{align}
where $\sigma_i$ describes the spin state at site $i$, and $A^{\sigma_i}$ is a matrix with dimension $\chi$.
The ground state is calculated by the imaginary-time evolution as 
\begin{align}
|\Psi_{\rm GS} \rangle \propto \prod^{N}_{n=1} \exp(-{\cal H} \delta t) |\Psi_{0}\rangle ,
\label{eq:iteb}
\end{align}
where $|\Psi_0 \rangle$ is the initial wave function, and $\delta t$ is small difference of the imaginary time. 
To calculate the exponential factor in Eq.~(\ref{eq:iteb}), we use the second order Suzuki-Trotter decomposition defined as 
$e^{A+B} \approx e^{A /2}e^{B}e^{A /2}. $
In the calculation of the time-evolved states, we also use Eq.~(\ref{eq:iteb}) where the imaginary time is replaced by the real time as $\delta t \rightarrow i \delta t$. 
We have confirmed that the numerical errors in the ground-state energies in $h_z/J \geq 1.1$ are less than $10^{-14}$ when $\chi = 50$. 
In the calculation of the real-time evolution, we adopt $\delta t = 0.01/J$ and 
the maximum number of $\chi$ is taken to be 200, by which the results are well converged. 
Relative numerical errors in $m^x$ are less than $10^{-2}$ for $h_z/J\leq 0.999$ and $10^{-5}$ for $h_z/J \leq 0.95$. 
From now on, we focus on the transient dynamics in the quantum disordered phase ($h_z/J > 1$). 
The pulse frequency is tuned at the energy gap, i.e. $\omega_{\mathrm{p}}={\mathit{\Delta}}$. 

\section{results}
\label{sec:results}
\subsection{Numerical Simulation}
\label{sec:itebd}
%
\begin{figure}[t]
\begin{center}
\includegraphics[width=1\columnwidth, clip]{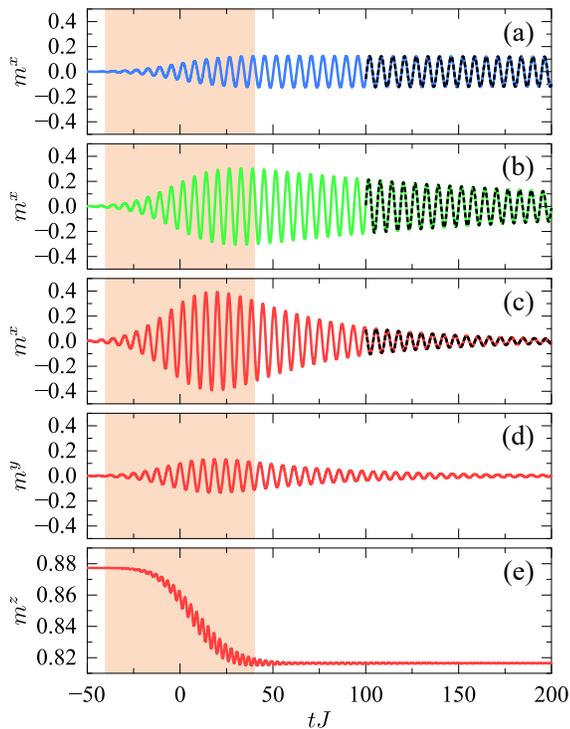}
\end{center}
\caption{(Color online) 
(a)-(c): Time profiles of the $x$ component of the magnetization induced by a short pulse field. 
Amplitude of the external field is (a) $A/J=0.001$, (b) 0.003, and (c) 0.005. 
Time profiles of the $y$ and $z$ components of the magnetization in the case of $A/J=0.005$ are shown in (d) and (e), respectively. 
Other parameter values are chosen to be $h_z/J=1.5$, $t_{\rm width}=0$, $\omega_{\mathrm{p}}/J=1$, and $t_{\mathrm{p}}=20/J$. 
Shaded areas represents the time domain, when the pump pulses are irradiated. 
Bold dashed lines in (a), (b), and (c) represent fitting curves for a time interval $t=100/J\text{--}200/J$.
}
\label{fig:sx_pulse}
\end{figure}

In this section, we show the numerical results obtained by the iTEBD algorithm. 
We first show the results under the short pulse field 
in which $t_{\rm width}=0$ in Eq.~(\ref{eq:field2}). 
Time profiles of $m^x(t)$ for several pulse amplitudes are shown in Figs.~\ref{fig:sx_pulse}(a)-(c). 
A quantum disordered state at $h_z/J=1.5$ is chosen as the initial state, in which a value of the energy gap is $\mathit{\Delta}/J=1$. 
In all cases, $m^x(t)$ begins to oscillate by the pulse field. 
The oscillation amplitude increases with increasing $A$. 
After turning off the pulse field, 
the oscillation in $m^x(t)$ remains with almost the same amplitude in $A/J=0.001$, while damping of the oscillation occurs in the cases of strong field, $A/J=0.003$ and $0.005$.
Dampings in the oscillations are also seen in $\sigma^y$ and $\sigma^z$ in the case of strong field as shown in Figs.~\ref{fig:sx_pulse}(d) and \ref{fig:sx_pulse}(e). 

\begin{figure}[t]
\begin{center}
\includegraphics[width=1\columnwidth, clip]{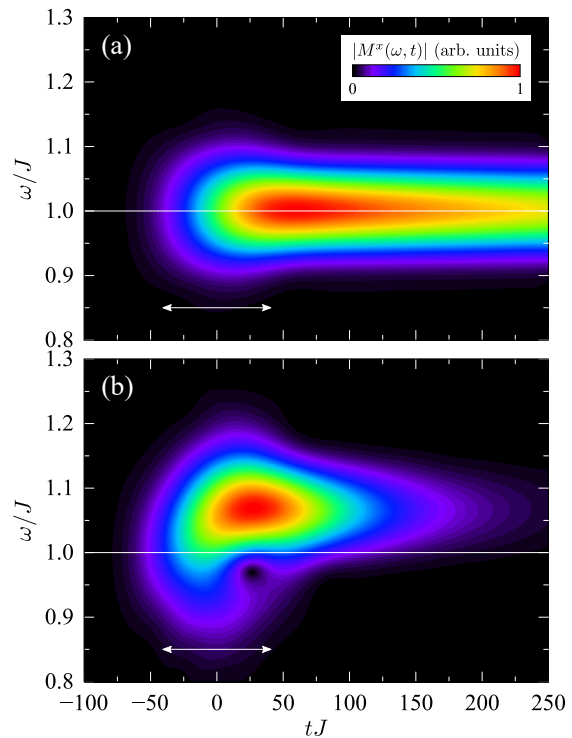}
\end{center}
\caption{(Color online) 
Contour maps of the short-time Fourier transformation of $m^x$ in the case of the short pulse field. 
Amplitudes of the pulse external field are chosen to be (a) $A/J=$0.002 and (b) 0.01. 
Other parameter values are chosen to be
$h_z/J=1.5$, $t_{\rm width}=0$, $\omega_{\mathrm{p}}/J=1$, and $t_{\mathrm{p}}=20/J$. 
Arrows represent the time domain ($-2t_{\rm p}<t < 2t_{\rm p}$), when the pump pulses are irradiated. 
In the short-time Fourier transformation, the Gaussian width is chosen to be $\tau = 24/J$. 
}
\label{fig:map}
\end{figure}

Transient dynamics induced by the short pulse field are analyzed using the Fourier decomposition. 
We apply the short-time Fourier transformation to $m^x(t)$ defined by 
\begin{align}
M^x(\omega, t)=\int_{t_1}^{t_2} dt'\, \frac{1}{\tau} e^{-i \omega t'} e^{-(t'-t)^2/(2\tau^2)} 
m^x(t') , 
\end{align}
where the Gaussian window is adopted. 
We chose $\tau = 24/J \text{--}64/ J$, $t_1=-100/J$, and $t_2=300/J$. 
Contour maps of $M^x(\omega, t)$ in a $\omega$-$t$ plane are presented in Figs.~\ref{fig:map}(a) and \ref{fig:map}(b) for $A/J=0.002$ and $0.01$, respectively. 
From now on, the frequencies at which absolute values of $M^x(\omega, t)$ are remarkable at each time are denoted as $\widetilde \omega$. 
In the case of weak pulse field [Fig.~\ref{fig:map}(a)], $\widetilde \omega$ is almost close to $\mathit{\Delta}$, and high intensity at $\widetilde \omega$ remains until $t=200/J$. A frequency shift from $\mathit{\Delta}$ and an intensity damping after turning off the pulse field are clearly seen in the case of strong pulse field shown in Fig.~\ref{fig:map}(b). 
In this case, by the pulse-field irradiation ($-40/J<t<40/J$), $\widetilde \omega$ begins to increase from $\mathit{\Delta}$, and is fixed around $1.08\mathit{\Delta}$. 
After turning off the pulse field ($t>40/J$), this frequency shift remains. 
Oscillation intensity is damped and almost disappears at around $t=150/J$ in contrast to the results in Fig.~\ref{fig:map}(a). 

\begin{figure}[t]
\begin{center}
\includegraphics[width=1\columnwidth, clip]{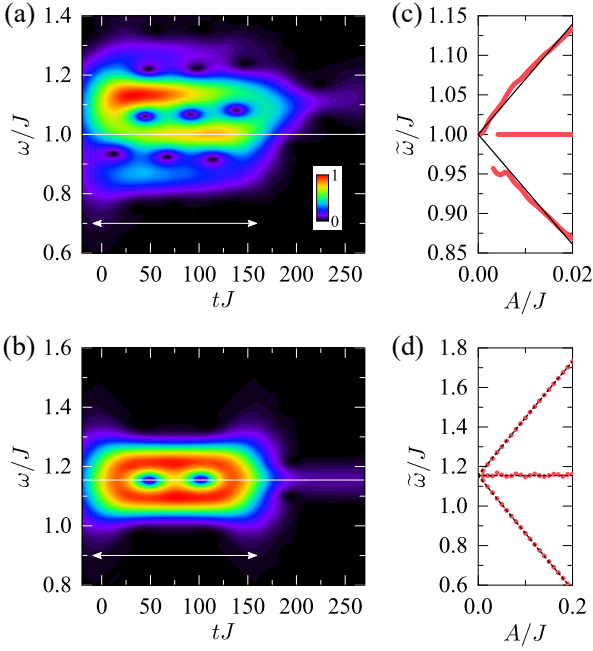}
\end{center}
\caption{(Color online) 
(a) A contour map of the short-time Fourier transformation of $m^x(t)$ obtained by the iTEBD algorithm in an infinite size system, and (b) a result obtained by the exact diagonalization method in a finite size cluster with $N=4$. 
Parameter values are chosen to be
$A/J=0.02$, $h_z/J=1.5$, $t_{\rm width}=150 /J$, $\omega_{\mathrm{p}}=\mathit{\Delta}$, and $t_{\mathrm{p}}=4 /J$. 
Arrows represent the time domains, when the pump pulses are applied. 
(c) Frequencies of the major oscillating components during the external-field irradiation ($0<t<t_{\rm width}$) obtained from the data in (a), and (d) the results obtained from the data in (b).
Numerical data in $M^x(\omega, t)$ are fitted by the three Gaussian functions. 
Bold lines in (c) are obtained by fitting, and dotted lines in (d) are the results by the Floquet theory presented in Sec.~\ref{sec:floquet}.
}
\label{fig:floquet2}
\end{figure}

In order to understand the dynamics more clearly, we show the results with the long pulse width. 
A contour map of $M^x(\omega, t)$ calculated by the iTEBD algorithm is presented in Fig.~\ref{fig:floquet2}(a). The pulse width is chosen to be $t_{\rm width}=150/J$. 
During the pulse irradiation ($0<t<150/J$), 
the frequency shifts from $\mathit{\Delta}\ (=J)$, and the intensity distribution shows asymmetric and broad structures.
After turning off the pulse field, the frequency shift from $\mathit{\Delta}$ is reduced slightly but remains. 
For comparison, we calculate a contour map of $M^x(\omega, t)$ in a $N=4$ size cluster by using the exact diagonalization method [see Fig.~\ref{fig:floquet2}(b)]. 
In contrast to the result in the infinite-size system in Fig.~\ref{fig:floquet2}(a), both the frequency shift and the intensity distribution are symmetric with respect to $\omega=\mathit{\Delta}$. 
After turning off the pulse field, the two frequencies are immediately merged into $\omega=\mathit{\Delta}$. 
Therefore, both the remaining of the frequency shift and the amplitude damping after turning off the pulse are attributable to the large cluster system. 

The characteristic oscillations in $m^x(t)$ during the pulse field irradiation are examined from the viewpoint of the Rabi oscillation~\cite{Bina2012} in the two-level system under the CW light. 
It is well known that the frequency of the Rabi oscillation is given by 
$\mathit{\Omega}_{\rm rabi}=\sqrt{(\mathit{\Delta}_{\rm two}-\omega_{\mathrm{p}})^2+A^2}$
, where $\mathit{\Delta}_{\rm two}$, $\omega_{\mathrm{p}}$, and $A$ are the energy difference between the two levels, the frequency, and the amplitude of the CW light, respectively. 
In the resonant case at $\omega_{\rm p}=\mathit{\Delta}_{\rm two}$, we have $\mathit{\Omega}_{\rm rabi}=\pm A$, which is proportional to the amplitude of the CW light. 
By fitting the numerical data shown in Figs.~\ref{fig:floquet2}(a) and (b) by the three Gauss functions, 
the major components of the oscillation frequencies during the pulse irradiation $(t=t_{\rm width}/2)$ are obtained. 
The results are plotted as functions of $A/J$ 
in Figs.~\ref{fig:floquet2}(c) and \ref{fig:floquet2}(d). 
It is shown that the frequencies are almost proportional to $A/J$ in both the two cases. 
Thus, the frequency shifts observed in Fig.~\ref{fig:floquet2} are considered to be a kind of the Rabi oscillation, in which the $\omega_{\rm p}$ is resonantly tuned at the gap energy $\mathit{\Delta}$.
Slopes of the frequency versus $A$ curves, denoted by $c_R$, are about $7$ and $2.5$ in Figs.~\ref{fig:floquet2}(c) and \ref{fig:floquet2}(d), respectively, which deviates from $1$ in the standard Rabi oscillation. 
We will examine $c_R$ in Sec.~\ref{sec:floquet}, and discuss the intersite exchange interaction effect.

\begin{figure}[t]
\begin{center}
\includegraphics[width=1\columnwidth, clip]{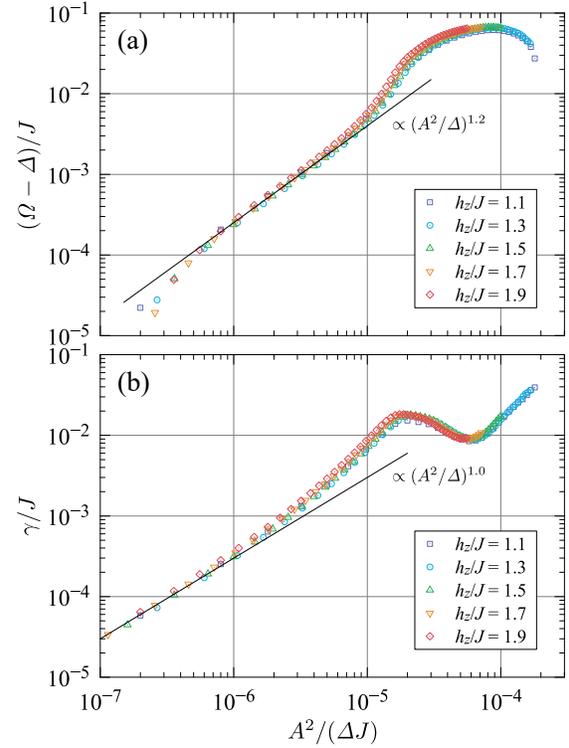}
\end{center}
\caption{(Color online) 
(a) Oscillation frequency $\mathit{\Omega}$ subtracted by $\mathit{\Delta}$, and (b) damping constant $\gamma$ plotted as functions of $A^2/\mathit{\Delta}$ after turning off the pulse field.
Both $\mathit{\Omega}$ and $\gamma$ are obtained by fitting the numerical data after turning off the pulse field. Parameter values are taken to be $h_z/J=1.1\text{--}1.9$, $t_{\rm width}=0$, $\omega_{\mathrm{p}}=\mathit{\Delta}$, and $t_{\rm p}=20/J$.
Solid lines represent $(A^2/\mathit{\Delta})^\alpha$ with $\alpha=1.2$ in (a) and $1.0$ in (b). 
}
\label{fig:scaling}
\end{figure}
%
Next, we focus on the transient dynamics after turning off the pulse field. 
To analyze the oscillations and damping, the numerical data shown in Fig.~\ref{fig:sx_pulse} are fitted by a function given by 
$m^x(t)=F e^{-\gamma t} \cos(\mathit{\Omega} t + \phi)$
where $F$, $\mathit{\Omega}$, $\gamma$, and $\phi$ are the fitting parameters of the amplitude, frequency, damping, and phase, respectively. 
Numerical fitting works well in the region after turning off the pulse field, as shown in Fig.~\ref{fig:sx_pulse}(c), where we have $F=0.598$, $\mathit{\Omega}=1.032J$, $\gamma=1.71 \times 10^{-2}J$, and $\phi=-2.05$. 
The calculated numerical data of $m^x(t)$ are fitted by this function for several values of $h_z/J\ (=1.1\text{--}1.9)$ and $A$. The obtained frequencies and damping factors are plotted in Figs.~\ref{fig:scaling}(a) and \ref{fig:scaling}(b), respectively, as functions of $A^2/\mathit{\Delta}$. 
Almost all data with different $h_z/J$ over four digits are located on single curves. 
Both $\mathit{\Omega}-\mathit{\Delta}$ and $\gamma$ monotonically increase with increasing $A^2/\mathit{\Delta}$ up to around $A^2/\mathit{\Delta}=10^5J$. 
In particular, data are well scaled by $(A^2/\mathit{\Delta})^\alpha$ with $\alpha=1.2$ for $\mathit{\Omega}$ in the region of $5\times10^7J <A^2/\mathit{\Delta} < 5 \times 10^{-4}J$, and $\alpha=1.0$ for $\gamma$ in $1\times 10^{-7}J < A^2/\mathit{\Delta} <1 \times 10^{-5}J$. 
These are plotted by solid lines in Fig.~\ref{fig:scaling}. 
That is, $\mathit{\Omega}-\mathit{\Delta}$ and $\gamma$ are almost proportional to $A^2/\mathit{\Delta}$. 
In the region of $A^2/\mathit{\Delta} < 5 \times 10^{-7}J$, 
numerical data for $\mathit{\Omega}$ are not scaled well by the universal line, since the limited time region is too short to estimate the frequency with good resolution. 
It is also found that energy increment from the ground-state energy due to the pulse external field is well fitted by $(A^2/\mathit{\Delta})^{1.0}$ in a region of $A^2/\mathit{\Delta} <10^{-5}J$ (not shown in figures).

\begin{figure}[t]
\begin{center}
\includegraphics[width=1\columnwidth, clip]{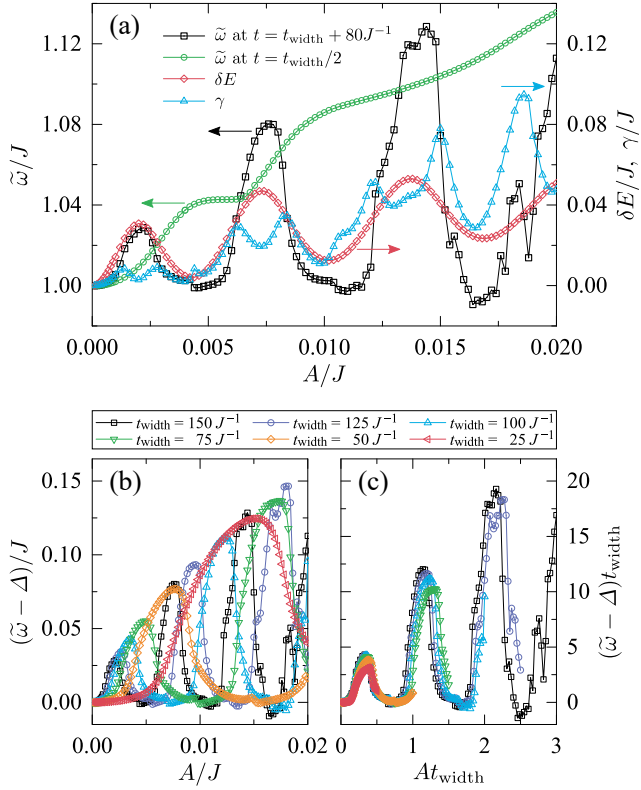}
\end{center}
\caption{(Color online) 
(a) Oscillation frequencies where $M^x(\omega, t)$ takes its maximum, termed $\widetilde{\mathit{\omega}}$, after turning off the pulse field $(t=t_{\rm width}+80/J)$, and those during the pulse irradiation $(t=t_{\rm width}/2)$ plotted by bold line with circles and bold line with squares, respectively. 
The energy increment $\delta E$ and the damping constant $\gamma$ are also plotted.
Width of the pulse fields are taken to be $t_{\rm width}=150/J$.
Oscillation frequencies are plotted as functions of $A/J$ in (b), and those as functions of $A t_{\rm width}$ in (c).
Parameter values are chosen to be $h_z/J = 1.5$, $\omega_{\rm p} = \mathit{\Delta} = 1.0J$, and $t_{\rm p}=4/J$.
}
\label{fig:cross}
\end{figure}

We examine the dynamics up to high fluence. 
In Fig.~\ref{fig:cross}(a), the characteristic oscillation frequency, $\widetilde{\mathit{\omega}}-\mathit{\Delta}$,
after turning off the pulse field ($t=t_{\rm width}+80/J$) up to $A/J=0.02$ are plotted by the bold line with squares.
We note that $\widetilde{\omega}$ were introduced as the frequencies in which $M^x(\omega, t)$ are remarkable at each time. Here we focus on $\widetilde{\omega}$ which is larger than $\mathit{\Delta}$.
The pulse width is chosen to be $t_{\rm width}=150/J$.
Note that numerical data presented in Fig.~\ref{fig:scaling} correspond to the region up to around $A/J=0.002$ in Fig.~\ref{fig:cross}(a), since $t_{\rm width}$ is chosen to be $0$ in Fig.~\ref{fig:scaling}.
After turning off the pulse field, large oscillations as function of $A/J$ are seen. 
%
For comparison, $\widetilde{\mathit{\omega}}-\mathit{\Delta}$ during the pulse irradiation $(t=t_{\rm width}/2)$ are also by the bold line with circles.
The frequency shift during the pulse irradiation increases monotonically with increasing $A/J$. 
It seems that the upper and lower bounds of $\widetilde{\omega}/J$ in $t>t_{\rm width}$ are given by 
$\widetilde{\omega}/J$ in $t<t_{\rm width}$ and $\mathit{\Delta}/J$, respectively.

The frequency shifts after turning off the pulse are shown in the Fig.~\ref{fig:cross}(b) for several values of $t_{\rm width}$. 
Periodicities of the oscillations strongly depend on $t_{\rm width}$. 
These data are replotted as function of $A t_{\rm width}$ in Fig.~\ref{fig:cross}(c). 
All data are well located on a single function in which the periodicity is deduced to be about $A t_{\rm width}=0.9$. 
This value is related to the frequency shift during the pulse irradiation shown in Fig.~\ref{fig:floquet2}(c); 
a time dependence of $m^x(t) $ follows approximately a function $\exp[i(\mathit{\Delta} \pm c_RA)t]$, 
where $c_R \sim 7$, corresponding to a periodicity $2\pi/c_R \sim 0.9$ as a function of $At$.
This scaling result implies that the oscillation after turning off the pulse is decided 
by the population of the excited state at around $t=t_{\rm width}$, i.e. the time when the pulse is turned off. 
This interpretation is supported by the results of the energy increment $(\delta E)$ from the ground state 
shown in Fig.~\ref{fig:cross}(a). 
The oscillation in $\delta E$ shows the same periodicity with $\widetilde \omega$ after turning off the pulse field (the bold line with squares in Fig.~\ref{fig:cross}),
indicating that a population of the excited state oscillates with the same frequency with $\widetilde \omega$.
Finally, we focus on the damping constants after turning off the pulse field. 
The results are presented in Fig.~\ref{fig:cross}(a).
It is found that $\gamma$ shows the local maxima at $A/J$ where the $\widetilde \omega$ versus $A/J$ curve is steep.
This fact implies that $\gamma$ is also decided by the population of the excited state when the pulse is turned off. 
In other words, the oscillations are stable, when the population of the excited state is minimum or maximum, 
but are damped largely, when the populations in the ground state and excited state are comparable. 

\subsection{Floquet Theory}
\label{sec:floquet}
%
\begin{figure}[t]
\begin{center}
\includegraphics[width=1\columnwidth, clip]{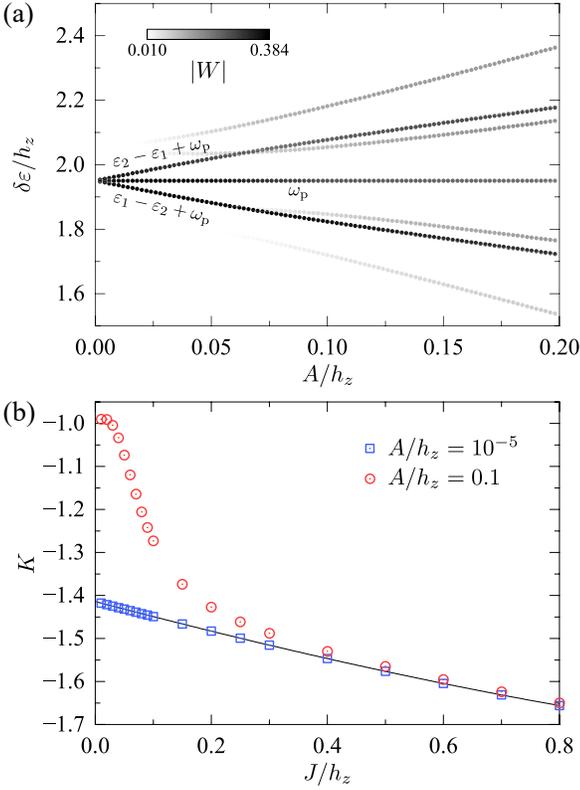}
\end{center}
\caption{(Color online) 
(a) Intensity plot of $|W|$ defined in Eq.~(\ref{eq:amplitude}) in a $\delta \varepsilon$-$A$ plane. 
A finite size cluster with $N=2$ is adopted. 
Other parameter values are chosen to be $J/h_z=0.05$ and $\omega_{\rm p}=\mathit{\Delta}$. 
(b) Slopes of the $\varepsilon_1-\varepsilon_2+\omega_{\rm p}$ versus $A$ curve in (a) at $A/h_z=10^{-5}$ and $0.1$. 
Bold line represents the proportional coefficient between $\varepsilon_1-\varepsilon_2$ and $A$ given in Eq.~(\ref{eq:AJ}).
}
\label{fig:slope}
\end{figure}

In this subsection, we examine the dynamics under the CW light by the Floquet theory,~\cite{Shirley1965, Sambe1973} in order to 
understand the numerical results during the pulse irradiation introduced in the previous section. 
We focus on the oscillation frequency in $\langle \sigma^x \rangle(t)$, and a proportional coefficient between this frequency and $A$ termed $c_R$.
We consider the TI model with the periodic external field, in which the envelop function $w_x(t)$ in Eq.~(\ref{eq:field}) is replaced by $1$, and $\omega_{\mathrm{p}}$ is chosen to be $\mathit{\Delta}$.

The time-dependent wave function is represented by a linear combination of the 
Floquet states given as 
\begin{align}
|\Psi(t) \rangle=\sum_\alpha c_\alpha e^{-i \varepsilon_\alpha t} \lvert \phi_\alpha(t)\rangle , 
\end{align}
where 
$|\phi_\alpha \rangle$ is the $\alpha$th Floquet state, $\varepsilon_\alpha$ is the corresponding quasi energy, and $c_\alpha$ is a complex number. 
By introducing the Fourier transformation defined by 
\begin{align}
|\phi_\alpha (t) \rangle=\sum_m e^{-im\omega_{\rm p} t}|\phi_\alpha^m \rangle ,
\end{align} 
the eigenvalue equation is obtained as 
\begin{align}
\sum_m \left ( {\cal H}_{n-m} -m \omega_{\rm p} \delta_{mn} \right ) |\phi_\alpha^m \rangle =\varepsilon_\alpha |\phi_\alpha^n \rangle . 
\label{eq:floquett} 
\end{align}
Here, ${\cal H}_m$ is the Fourier transformation of the time-dependent Hamiltonian ${\cal H}(t)$ defined by 
\begin{align}
\mathcal{H}_m = \frac{\omega_{\rm p}}{2\pi} \int_0^{2\pi/\omega_{\rm p}} dt\, e^{im\omega_{\rm p} t} \mathcal{H}(t).
\end{align}
In the present model, we have
\begin{align}
\mathcal{H}_0 = \mathcal{H}_{\rm TI},
\end{align}
\begin{align}
\mathcal{H}_{\pm 1}&= - \frac{A}{2} \sum_i \sigma_i^x ,
\end{align}
and $\mathcal{H}_{\pm n} = 0$ for $n\geq 2$. 
The expectation value of $\sigma^x$ at time $t$ is obtained by 
\begin{align}
\langle \sigma^x \rangle(t)&=\langle \Psi(t) |\sigma^x|\Psi(t) \rangle \nonumber \\
&= \sum_{m n \alpha \beta} e^{i \delta \varepsilon t} W ,
\label{eq:flq}
\end{align}
where we define the frequency difference 
\begin{align}
\delta \varepsilon =(\varepsilon_\beta-\varepsilon_\alpha)+(m-n) \omega_{\rm p} , 
\label{eq:epsilon}
\end{align}
and the weight
\begin{align}
W = c_\alpha^\ast c_\beta \langle \phi_\alpha^m | \sigma^x | \phi_\beta^n \rangle . 
\label{eq:amplitude}
\end{align} 

In the framework of the Floquet theory, 
we analyze the dynamics during the pulse irradiation shown in Figs.~\ref{fig:floquet2}(b) and \ref{fig:floquet2}(d). 
The eigenvalue equation in Eq.~(\ref{eq:floquett}) is solved in a finite size cluster with $N=4$. We consider the Floquet states up to the two-photon dressed states, corresponding to that the upper and lower limits of the summation in Eq.~(\ref{eq:floquett}) are taken to be $2$ and $-2$, respectively. 
The two quasi-energies, which are bound for the ground state of ${\cal H}_0$ and the first excited state of ${\cal H}_0-\omega_{\rm p}$ in the limit of $A \rightarrow 0$, are identified as $\varepsilon_1$ and $\varepsilon_2$. 
These are plotted as a function of $A$ in Fig.~\ref{fig:floquet2}(d) by dotted lines. 
The results by the Floquet theory well reproduce the numerical data obtained by the exact diagonalization method in a cluster with $N=4$. 
That is, the frequency shifts during the pulse irradiation 
are understood in the Floquet theory. 

We also analyze $c_R$, the proportional coefficient between the oscillation frequencies in $\langle \sigma^x \rangle(t)$ and $A$. 
We solve the eigenvalue equation in Eq.~(\ref{eq:floquett}) in a small cluster with $N=2$.
In Fig.~\ref{fig:slope}(a), we present an intensity plot of $W$ in a $\delta \varepsilon$-$A$ plane, in which the $m-n=1$ component contributes. 
As shown in the figure, the three states provide the major contributions to $\langle \sigma^x \rangle(t)$. 
These are identified as $\varepsilon_2-\varepsilon_1+\omega_{\rm p}$, 
$\varepsilon_1-\varepsilon_2+\omega_{\rm p}$, and 
$\omega_{\rm p}$, which correspond to the three frequencies in Fig.~\ref{fig:floquet2}(d).
%
We evaluate slopes of the 
$\varepsilon_1-\varepsilon_2+\omega_{\rm p}$ versus $A$ curve in Fig.~\ref{fig:slope}(a), 
defined by
\begin{align}
K=\frac{\partial }{\partial A} \delta \varepsilon,
\label{eq:slope}
\end{align}
and plot the results as functions of the exchange interaction $J/h_z$ in Fig.~\ref{fig:slope}(b). 
Absolute values of the slopes are strongly enhanced by the exchange interaction. 
The slopes are varied by $A$, but approach asymptotically to a single line for large $J$. 
In the case of $A/h_z=0.1$, we confirm that $|K|=1$ at $J \rightarrow 0$ as expected in the standard Rabi oscillation, and the crossover occurs around $J \sim A$. 

In order to understand the above results in more detail, we analyze the two relevant Floquet states, 
$|\phi_1\rangle$ and $|\phi_2 \rangle$, 
which are bound for the ground state of ${\cal H}_0$ and the first excited state of ${\cal H}_0-\omega_{\rm p}$ in the limit of $A \rightarrow 0$, respectively.
The two states are degenerated at $A=0$, 
and the degeneracy is lifted by the first-order perturbation of $\mathcal{H}_{\pm 1}$.
By diagonalizing the $2\times 2$ matrix obtained by the first-order perturbation given by 
\begin{align}
\begin{pmatrix}
0 &&
\langle \phi_1 \vert \mathcal{H}_{- 1} \vert \phi_2 \rangle \\
\langle \phi_2 \vert \mathcal{H}_{+ 1} \vert \phi_1 \rangle &&
0
\end{pmatrix} , 
\label{eq:matrix}
\end{align}
the quasi energies are obtained as
\begin{align}
\varepsilon_{1,2}= 
C \pm
\frac{A}{\sqrt{2}}
\sqrt{1+
\frac{J}{\sqrt{4h_z{}^2+J^2}}
} . 
\label{eq:AJ}
\end{align}
A constant $C$ 
is independent of $A$. 
We plot the proportional coefficient between $\varepsilon_1-\varepsilon_2$ and $A$ by a bold line in Fig.~\ref{fig:slope}(b). 
This gives the asymptotic line of $K$ in the region of large $J$. 
Enhancement of $c_R$ from $1$ in the standard Rabi oscillation is seen in 
the region of $A \ll J, h_z$ and is attributed to the exchange interaction effects. 
Physical picture of this enhancement is due to the off-diagonal matrix elements in Eq.~(\ref{eq:matrix}). 
The wave functions are approximately given by 
\begin{align}
|\phi_1\rangle=c_1 \nobreak{\lvert\uparrow, \uparrow \rangle} +c_2 \nobreak{\lvert\downarrow , \downarrow \rangle}
\end{align}
and 
\begin{align}
|\phi_2\rangle=\frac{1}{\sqrt{2}} \left(\nobreak{\lvert\uparrow, \downarrow \rangle} + \nobreak{\lvert\downarrow, \uparrow\rangle} \right), 
\end{align}
where $|\sigma_1, \sigma_2\rangle $ represents the spin states at the two sites.
The coefficients are given as $(c_1, c_2)=(1, 0)$ for $J=0$ and $(1/\sqrt{2}, 1/\sqrt{2})$ for $J \rightarrow \infty$. 
The matrix element $\langle \phi_2 | {\cal H}_{+1} | \phi_1 \rangle$ is changed from $-A/\sqrt{2}$ to $-A$ with increasing $J$ from $0$, implying increasing of the mixing channel between the zero and one-photon states by the photon absorption. 

\section{summary}
\label{sec:discussion}

We study the transient dynamics of a prototypical low-dimensional interacting quantum system, i.e. 
the one-dimensional TI model, induced by the oscillating pulse excitation and their connection to the dynamics by the CW excitation. 
This model is relevant for several physical systems, and has some similarities to the studies in interacting cold atom systems.~\cite{Cummings1983, Lukin2001, Dudin2012}
We stress that the present numerical results using the iTEBD algorithm enable us to obtain the unambiguous results owing to the interaction effects without a finite-size effect.
We focus on (i) the time domain during the pulse irradiation $(t<t_{\rm width})$, and (ii) that after turning off the pulse $(t>t_{\rm width})$. 
In the time domain (i), the characteristic oscillation in $\langle \sigma^x \rangle (t)$ is understood in the Floquet theory as a generalized Rabi oscillation. 
The oscillation frequency shifts from the standard Rabi oscillation, and its proportional coefficient to $A$, termed $c_R$, is enhanced by the intersite exchange interaction. 
A value of $c_R$ is smoothly changed to $1$ when $J$ is smaller than $A$.
In the time domain (ii), the frequency shift from the pulse frequency remains and the amplitude of the oscillation is damped. 
In the case of low fluence, the frequency shift is well scaled by $A^2/\mathit{\Delta}$ in a wide parameter region of $h_z/J$, even away from the quantum critical point of $h_z/J=1$. 
In the case of high fluence, on the other hand, both the frequency shift and the damping factor strongly depend on $t_{\rm width}$ and $A$, and is well arranged by the product, $A t_{\rm width}$. 
This observation implies that the characteristic dynamics in this time domain are decided by the population of the photoexcited state when the pulse field is turned off. 
Finally, we briefly comment on the experimental feasibility.
In the present numerical calculations, amplitude of the pulse field is chosen to be up to of the order of $A/J = 0.01$. 
This corresponds to $A \sim 150\,{\rm kV/cm}/c$ with the light velocity $c$,
in which we assume $J \sim 20\,{\rm K}$ and the critical transverse field $h_z^{\rm c} \sim 5\,{\rm T}$ which are reasonable values for realistic materials.~\cite{Kinross2014, Coldea2010}
This value is experimentally accessible by the recent THz laser techniques.

 \begin{acknowledgments}
The authors would like to thank M. Naka and S. Iwai for valuable discussions. 
This work was supported by MEXT KAKENHI Grant Numbers 26287070 and 15H02100. 
Some of the numerical calculations were performed using the facilities of the Supercomputer Center, the Institute for Solid State Physics, the University of Tokyo.
 \end{acknowledgments}

\bibliography{reference}
\end{document}